\documentclass[10pt,twocolumn,floatfix,superscriptaddress,amsmath,showkeys,aps,prb,eqsecnum]{revtex4-2}
\usepackage{bm}
\usepackage{xcolor}
\usepackage{hyperref}
\usepackage{makeidx}
\usepackage{amsmath}
\usepackage{graphicx}
\usepackage{dcolumn}
\usepackage{float}
\usepackage{amsfonts}
\usepackage{amssymb}
\usepackage{color}
\usepackage{array}
\usepackage{soul}

\newcommand {\vk} {\mathbf{k}}
\newcommand {\vQ} {\mathbf{Q}}

\newcommand {\vR} {\mathbf{R}}
\newcommand \sa {\sigma}
\newcommand \up {\uparrow}
\newcommand \pu {\downarrow}
\newcommand \epk[1] {\epsilon_{{\bf k}}^{#1}}

\newcommand \epks[2] {\epsilon_{{\bf k}{#2}}^{#1}}
\newcommand \epkqs[2] {\epsilon_{{\bf k+Q}{#2}}^{#1}}
\newcommand \epkms[2] {\epsilon_{{\bf -k}{#2}}^{#1}}
\newcommand \epkqms[2] {\epsilon_{{\bf -k-Q}{#2}}^{#1}}

\newcommand \epd {\epsilon_{d0}}
\newcommand \ckm[1]  {c_{#1}^{\dag}}
\newcommand \ck[1]     {c_{#1}}
\newcommand \fkm[1] {d_{#1}^{\dag}}
\newcommand \fk[1] {d_{#1}}
\newcommand \med[1] {\langle{#1}\rangle}
\newcommand \ntot {n_{\mathrm{tot}}}

\hypersetup{
pdfauthor = {Lopes, Reyes, Costa, Continentino, and Thomas},
pdftitle = {InCDW on multiband intermetallic systems exhibiting competing orders},
}

\begin{document}

\title{Incommensurate charge density wave on multiband intermetallic systems exhibiting competing orders}

\author{Nei Lopes} 
\email{nlsjunior12@gmail.com}
\affiliation{Departamento de F\'{\i}sica Te\'orica, Universidade do Estado do Rio de Janeiro, Rua S\~ao Francisco Xavier 524, Maracan\~a, 20550-013, Rio de Janeiro, RJ, Brazil}
\author{Daniel Reyes} 
\affiliation{Instituto  Militar  de  Engenharia -  Pra\c{c}a  General Tib\'urcio 80, 22290-270, Praia Vermelha, Rio de Janeiro, Brazil}
\affiliation{Laboratorio de Cer\'amicos y Nanomateriales, Facultad de Ciencias F\'{\i}sicas, Universidad Nacional Mayor de San Marcos, Ap. Postal 14-0149, Lima, Peru}
\author{Natanael C. Costa}
\affiliation{Instituto de F\'isica, Universidade Federal do Rio de Janeiro, Rio de Janeiro, RJ 21941-972, Brazil}
\author{Mucio A. Continentino} 
\affiliation{Centro Brasileiro de Pesquisas F\'isicas, Rua Dr.~Xavier Sigaud 150, Urca, 22290-180, Rio de Janeiro, Brazil}
\author{Christopher Thomas} 
\affiliation{Instituto de F\'isica,  Universidade Federal do Rio Grande do Sul, 91501-970, Porto Alegre, Brazil}

%%%%%%%%%%%%%%%%%%%%%%%%%%%%%%%%%%%%%%%%%%%%%%%%%%%%
\begin{abstract}
The appearance of an incommensurate charge density wave vector $\vQ = (Q_x,Q_y)$ on multiband intermetallic systems presenting commensurate charge density wave (CDW) and superconductivity (SC) orders is investigated. We consider a two-band model in a square lattice, where the bands have distinct effective masses. The incommensurate CDW (inCDW) and CDW phases arise from an interband Coulomb repulsive interaction, while the SC emerges due to a local intraband attractive interaction. For simplicity, all the interactions, the order parameters and hybridization between bands are considered $\vk$-independent. The multiband systems that we are interested are intermetallic systems with a $d$-band coexisting with a large $c$-band, for which a mean-field approach has proved suitable. We obtain the eigenvalues and eigenvectors of the Hamiltonian numerically and minimize the free energy density with respect to the diverse parameters of the model by means of the Hellmann-Feynman theorem. We investigate the system in real as well as momentum space and we find an inCDW phase with wave vector $\vQ = (\pi, Q_y) = (Q_x, \pi)$. Our numerical results show that the arising of an inCDW state depends on parameters, such as the magnitude of the inCDW and CDW interactions, band filling, hybridization and the relative depth of the bands. In general, inCDW tends to emerge at  low temperatures, away from half-filling. We also show that, whether the CDW ordering is commensurate or incommensurate, large values of the relative depth between bands may suppress it. We discuss how each parameter of the model affects the emergence of an inCDW phase.
\end{abstract}
%%%%%%%%%%%%%%%%%%%%%%%%%%%%%%%%%%%%%%%%%%%%%%%%%%%%

\date{\today}
\maketitle

%%%%%%%%%%%%%%%%%%%%%%%%%%%%%%%%%%%%%%%%%%%%%%%%%%%%%%%%%%
\section{Introduction}
\label{sec:intro}
%%%%%%%%%%%%%%%%%%%%%%%%%%%%%%%%%%%%%%%%%%%%%%%%%%%%%%%%%%%

The search for coexistence, competition, or even a cooperative behavior between superconductivity (SC) and other collective states, such as magnetism and charge density wave (CDW)  may shed light on novel and exotic states of matter. However, understanding the emergence and interplay of these different types of electronic order is a challenge, and remains an ongoing topic of research. Prominent examples include SC in high-temperature copper oxide superconductors~\cite{Bussmann-Holder1992,Birgeneau1987,Wakimoto2004}, organic charge-transfer salts~\cite{Gabovich2001,Lubczynski1996,Wosnitza2001}, heavy-fermion
compounds~\cite{Grosche2001}, superconducting cobalt systems~\cite{Takada2003}, A15 compounds~\cite{Chu1974,Chu1974a,Testardi1975,Tanaka2010}, Ni- and Fe-based superconductors~\cite{delaCruz2008,Yoshizawa2012,Niedziela2011,Kudo2012,Hirai2012}, perovskites~\cite{Kang2011}, quasi-skutterudite superconductor~\cite{Kase2011, Wang2012,Klintberg2012,Zhou2012,Biswas2014,Goh2015}, intercalated graphite CaC$_6$~\cite{Gauzzi2007},  sulfuride-based compounds at very high pressure~\cite{Degtyareva2007},
and some transition-metal dichalcogenides (TMDs)~\cite{Gabovich2001,Morosan2006,Zhao2007,Wilson1975,Kusmartseva2009,Sipos2008,Yang2012,Pyon2012,Fang2013,Kamitani2016,Kudo2016,Heil2017,Saint-Paul2021, Friend1987,Manzeli17,Chua2022,multiCDW2021}.

Within this context, TMD materials have gained widespread attention due to parallels and similarities between their electronic phases and/or interactions with those observed in high-temperature superconducting copper oxide and iron arsenide materials~\cite{Rossnagel2011,Borisenko2009,Chatterjee15}.
The occurrence of \textit{strange metal} behavior, Mott insulating phases, pseudogap states~\cite{Borisenko2009,Chatterjee15} and  the emergence of two superconducting domes in the phase diagram of $1T$-Cu$_x$TiSe$_2$, as a function of Cu intercalation or pressure, in proximity of the CDW state~\cite{Kusmartseva2009} have pointed out analogies between these compounds. It has also intensified the debate about the role of the electron-electron and electron-phonon interactions~\cite{Chua2022,Rossnagel2011}. Of particular interest is the response of such materials to impurity dopants or pressure, which are known to tune the structural and electronic properties of TMD materials~\cite{Wu1990,Wu1988,Bando2000,Su2012,Araujo22,SousaJr23}. 

In principle, the SC observed in these compounds can be investigated within the framework of Bardeen-Cooper-Schrieffer (BCS) theory~\cite{Bardeen1957,Bardeen1957a} due to the nodeless nature of the superconducting gap function~\cite{Kase2011,Hayamizu2011,Wang2012,Zhou2012,Biswas2014}. Furthermore, this is supported by the temperature dependence of the specific heat, and the ratios $2\Delta/k_B T_{\mathrm{SC}}$ and $\Delta C/\gamma T_{\mathrm{SC}}$ close to the expected values of the BCS theory~\cite{Bardeen1957,Bardeen1957a}. However, for others materials/compounds it  is necessary to go well beyond the BCS picture. This is the case when the pairing mechanism may not be phonon mediated, and the SC becomes unconventional, i.e., being enhanced by quantum fluctuations and accompanied by the suppression of the CDW to a quantum critical point (QCP)~\cite{Lee2021,Gruner2017}.

In addition, it has been observed that several TMDs exhibit an interplay between SC and incommensurate charge density wave (inCDW) orders~\cite{Sipos2008,Liu2013,Li2017,Gabovich2002,Wagner2008,Kogar2017,Chen2015,Wen2020,Song2022,Flicker16,Bhoi2016}, while the coexistence of SC and commensurate CDW [hereafter denoted only by CDW] is relatively rare~\cite{Yokota2000}. Recently, it has been shown that coexistence  depends directly on the  band filling and the relative depth of the bands~\cite{Lopes2021}. This interplay has also been reported in Ni- and Fe-based pnictides~\cite{Lee2019,Sefat2009}, and in Y-, Bi-, and Hg-based  cuprates~\cite{Ghiringhelli2012,Chang2012,Achkar2012,Comin2014,daSilvaNeto2014,Wu2015,Tabis2014,Blackburn2013,LeTacon2014} and rare-earth intermetallic systems~\cite{Carneiro2020}. Therefore, further studies on the relation between inCDW, CDW and SC states might lead to a deeper understanding of this collective quantum states in solids.

In order to contribute to such a discussion, we investigate the interplay of CDW, inCDW, and SC on intermetallic systems.
In particular, we examine how an incommensurate charge modulation wave vector $(\vQ)$, which give rises to an inCDW state, appears in the phase diagrams of multiband intermetallic systems, and how it affects the emergence of SC. To this end, we disregard the complexities behind specific compounds, and investigate their \textit{global} fundamental aspects by means of effective lattice Hamiltonians, focusing on the incommensurability features. Here, we consider a two-band model in a square lattice, whose bands exhibit different effective masses~\onlinecite{Lopes2021}. The inCDW and CDW phases arise from an interband Coulomb repulsive interaction, while SC emerges due to a local intraband attractive interaction. 

However, even dealing with simplified Hamiltonians, the analysis of inCDW is a challenge: unbiased methods, such as quantum Monte Carlo or density matrix renormalization group, may not be adequate, whether because large lattice sizes are too computationally demanding, or because of technical constraints (as the fermionic minus sign problem). Therefore, we analyze it through a mean-field theory in both real and momentum space configurations. Using a self-consistent procedure to minimize the free energy density, we are able to obtain phase diagrams for our model that exhibit a plethora of phases and examine their interplay and competition. We show that the appearance of an inCDW state depends on many parameters, such as temperature, band filling, hybridization, and on-site orbital energies.
For instance, a strong interband Coulomb interaction suppresses an inCDW order. This  incommensurate charge ordered phase tends to emerge in the low temperature regime, away from half-filling, and close to the transition between CDW and SC, and pure SC, in the absence of hybridization. By contrast, for large hybridization the inCDW appears near the half-filling.

It is worth to point out that our phase diagrams for CDW, inCDW, and SC orders are in line with those obtained from mainly phononic interactions\,\cite{Dee19}.
Despite having different natures, phononic and electronic models share some similarities: for instance, from a Lang-Firsov transformation, one is able to map electron-phonon systems in electronic ones, by integrating out the bosonic degrees of freedom, leading to non-local interactions\,\cite{Lang63,Hohenadler04}.
That is, fundamental properties of the competition between charge and pairing orders may be described with both approaches. We expect that, notwithstanding we are dealing with an electronic model, our study can give further insights on the emergence of CDW, inCDW and SC. 

The paper is organized as follows: In Sec.~\ref{sec:model} we describe the main aspects of our two-band model to investigate the effects of inCDW, CDW and SC orders on multiband intermetallic systems as well as the mean-field approximation used to solve the many-body problem. In Sec.~\ref{sec:results} we present our numerical results in real and momentum space, focusing in the phase diagrams as a function of different parameters. In Sec.~\ref{sec:concl} we conclude and make some remarks about our main results.

%%%%%%%%%%%%%%%%%%%%%%%%%%%%%%%%%%%%%%%%%%%%%%
\section{Model and methods}
\label{sec:model}
%%%%%%%%%%%%%%%%%%%%%%%%%%%%%%%%%%%%%%%%%%%%%%%%
%%%%%%%%%%%%%%%%%%%%%%%%%%%%%%%%%%%%%%%%%%%%%%
\subsection{The model}
%%%%%%%%%%%%%%%%%%%%%%%%%%%%%%%%%%%%%%%%
We consider a two-band model consisting of a large $c$-type band and a narrower one with moderate correlations, in a square lattice. The latter has essentially a $d$-character. We take into account on-site interband Coulomb repulsion between the bands, that give rises to CDW/inCDW, and local attractive intraband interactions in the $d$-band, which is responsible for SC.

The real-space Hamiltonian of the model reads~\onlinecite{Lopes2021}
\begin{align}\label{ref:hamiltonian}
H =&  - t_c\sum_{\langle {i} {j} \rangle, \sa} \big(\ckm{{i}\sa}\ck{{j}\sa} + {\rm H.c.} \big)
- t_d\sum_{\langle {i} {j} \rangle, \sa} \big( \fkm{{i}\sa}\fk{{j}\sa} \notag + {\rm H.c.} \big)\\
& + \epsilon_{d0}\sum_{{i}, \sigma} \fkm{{i}\sa}\fk{{i}\sa} - \mu  \sum_{{i}, \sigma} \big( \fkm{{i}\sa}\fk{{i}\sa} + \ckm{{i}\sa}\ck{{i}\sa}  \big) \notag \\
&+ \sum_{{i,j},\sa}V_{ij}(\ckm{{i}\sa}\fk{{j}\sa}+\fkm{{i}\sa}\ck{{j}\sa}) \notag\\
 &+U_{dc}\sum_{{i}}n^d_{{i}} n^c_{{i}}+ J_d\sum_{{i}} \fkm{{i}\up}\fk{{i}\up}\fkm{{i}\pu}\fk{{i}\pu} ~,%\notag \\
\end{align}
where $c_{{i}\sigma}$ ($c^{\dagger}_{{i}\sigma}$) and $d_{{i}\sigma}$ ($d^{\dagger}_{{i}\sigma}$) denote annihilation (creation) operators of $c$- and $d$-electrons, respectively, in a given site ${i}$, with spin $\sigma$, in the standard second quantization formalism.
The first two terms on the right-hand side of the Eq.\,\eqref{ref:hamiltonian} correspond to the hopping of $c$- and $d$-bands, with $t_{c,d}$ defining their hopping integrals. The third term defines the relative shift $\epsilon_{d0}$ between $c$- and $d$-bands, while the chemical potential $\mu$, in the fourth term, is self-consistently determined to accommodate a given total number $\ntot$ of electrons. The fifth one denotes the hybridization between the orbitals. For simplicity, we define the  hybridization $V_{ij}$ between different orbitals on neighboring sites as real and symmetric. The last two terms denote the on-site interband electronic repulsion~($U_{dc} > 0$), and an on-site effective attraction between $d$-electrons~($J_d < 0$).

Since our main interest is to investigate charge orderings,  we allow for the occurrence of commensurate and incommensurate charge density waves by means of a modulation of the average values of the occupation numbers~\cite{Brydon2005}, 
\begin{align}\label{eq:density1}
\med{n_{{i}}^c}&=n^c+\delta^c\cos{(\vQ\cdot\vR_i)}\,,\\
\label{eq:density2}
\med{n_{{i}}^d}&=n^d+\delta^d\cos{(\vQ\cdot\vR_i)}\,,
\end{align}
where $\delta^{c}$ and $\delta^{d}$ play the role of the CDW/inCDW order parameters for $c$- and $d$-orbitals, respectively, while $\vQ = (Q_x,Q_y)$ is the modulation wave vector~\cite{Chikina2020}. It is worth mentioning that, by incommensurate, we meant a wave vector different from $(\pi,\pi)$, with $Q_\alpha = \frac{2 \pi l}{L}$, $l = -\frac{L}{2}, \dots, \frac{L}{2} -1$, and $L$ being the linear size of the lattice.
Also, we define the number of particles for each band as $n^{d(c)}= n_{\up}^{d(c)}+n_{\pu}^{d(c)}$, and  disregard magnetic solutions by enforcing $\langle n_{\up}^{d(c)} \rangle = \langle n_{\pu}^{d(c)} \rangle$. We investigate the properties of the Hamiltonian in Eq.\,\eqref{ref:hamiltonian} by a static mean-field theory, performing this approach for real and momentum spaces, whose procedures are outlined in the next two subsections, respectively. 

%%%%%%%%%%%%%%%%%%%%%%%%%%%%%%%%%%%%%%%%%%%%
\subsection{The real-space mean-field approach}
%%%%%%%%%%%%%%%%%%%%%%%%%%%%%%%%%%%%%%%%%%%%%%%

By performing a Hartree-Fock approach on the interacting terms of Eq.~(\ref{ref:hamiltonian}), and using Eqs.\,\eqref{eq:density1} and \eqref{eq:density2} one obtains
\begin{align}\label{eq:realspaceHamil}
\nonumber
H_{\rm MF} & = 
-t_{c}\sum_{\langle i, j \rangle \sigma} \big( c^{\dagger}_{\mathbf{i}\sigma} c_{\mathbf{j}\sigma} + {\rm H.c.}\big) 
-t_{d}\sum_{\langle i, j \rangle \sigma} \big( d^{\dagger}_{\mathbf{i}\sigma} d_{\mathbf{j}\sigma} + {\rm H.c.}\big)
\\
\nonumber
& + V\sum_{ i \sigma} \big( c^{\dagger}_{\mathbf{i}\sigma} d_{\mathbf{i}\sigma} + {\rm H.c.}\big) 
+ \Delta_{d}\sum_{ i } \big( d^{\dagger}_{\mathbf{i}\uparrow} d^{\dagger}_{\mathbf{i}\downarrow} + {\rm H.c.}\big)
\\
\nonumber
& + \sum_{ i \sigma} \big[ -\mu + U_{dc} \big(n_d + \delta_d \cos(\mathbf{Q}\cdot\mathbf{R}_{i}) \big) \big] c^{\dagger}_{\mathbf{i}\sigma} c_{\mathbf{i}\sigma} 
\\
\nonumber
& + \sum_{ i \sigma} \big[ \epsilon_{d0} -\mu + U_{dc} \big( n_c + \delta_c \cos(\mathbf{Q}\cdot\mathbf{R}_{i}) \big) \big] d^{\dagger}_{\mathbf{i}\sigma} d_{\mathbf{i}\sigma} 
\\
& - N\bigg( \frac{\Delta^{2}_{d}}{J_d} + U_{dc} n_d n_c + N_{\mathbf{Q}} U_{dc} \delta_d \delta_c  -\mu (n_c + n_d)\bigg).
\end{align}
Here we assume a local hybridization (i.e., $V_{ij}= \delta_{ij} V$), while defining
\begin{equation}\label{eq:Deltasc_real}
\Delta_d = \frac{1}{N} \sum_{i} \langle d^{\dagger}_{\mathbf{i}\uparrow} d^{\dagger}_{\mathbf{i}\downarrow} \rangle
= \frac{1}{N} \sum_{i} \langle d_{\mathbf{i}\downarrow} d_{\mathbf{i}\uparrow} \rangle~,
\end{equation}
and
\begin{equation}
N_{\mathbf{Q}}= \left \{
    \begin{array}{c l}	
         1 & {\rm if\,\,\mathbf{Q}=}\,\,(\pm\pi,\pm\pi); \\
         1/2 & {\rm otherwise}.
    \end{array}\right.
\end{equation}

Notice that the Hamiltonian in Eq.\eqref{eq:realspaceHamil} may be written in a $4N \times 4N$ matrix representation in a basis $\{c^{\dagger}_{\uparrow} c_{\downarrow} d^{\dagger}_{\uparrow} d_{\downarrow} \}$,
with $N=L\times L$ being the number of sites.
That is, the Nambu spinor may be defined as $\Psi^{\dagger} = \big( c^{\dagger}_{1 \uparrow}, \dots, c^{\dagger}_{N \uparrow}, c_{1 \downarrow}, \dots, c_{N \downarrow}, d^{\dagger}_{1 \uparrow}, \dots, d^{\dagger}_{N \uparrow}, d_{1 \downarrow}, \dots, d_{N \downarrow}\big)$.

The diagonalization of $H_{MF}$ provides $4N$ eigenvalues $\lambda_{i}$ which allow us to obtain the free energy density,
\begin{align}\label{free_energy}
F= -\frac{T}{N} \sum_{i} \ln{[1+\exp{(-\beta \lambda_{i})}]} + const.,\
\end{align}
where $\beta=1/(k_B T)$, with $k_B$ being the Boltzmann constant, and $T$ the absolute temperature.
The numerical minimization of the free energy density with respect to the mean-field parameters, i.e.
\begin{equation}\label{self-consistent_eqs}
\frac{\partial F}{\partial \mu} = \frac{\partial F}{\partial n^d} = \frac{\partial F}{\partial \delta^d} = \frac{\partial F}{\partial \delta^c} = \frac{\partial F}{\partial \Delta^d} = \frac{\partial F}{\partial Q_\alpha} = 0,
\end{equation}
is performed self-consistently with the aid of the Hellmann-Feynman theorem~\cite{Feynman39,Stanton62,Jensen07}.

%%%%%%%%%%%%%%%%%%%%%%%%%%%%%%%%%%%%%%%%%%%%%%%%%%%%%%%%%%%%%%%%%%%%%%%%%%
\subsection{The momentum-space mean-field approach}\label{subsec:mspace}
%%%%%%%%%%%%%%%%%%%%%%%%%%%%%%%%%%%%%%%%%%%%%%%%%%%%%%%%%%%%%%%%%%%%%%%%%%
By performing a Fourier transform of the mean-field Hamiltonian in Eq.\,\eqref{eq:realspaceHamil}, one obtains
\begin{align}\label{eq:momentumspaceHamil}
\nonumber
H_{\rm MF} & = 
\sum_{\mathbf{k} \sigma} \epsilon^{c}_{\mathbf{k}} c^{\dagger}_{\mathbf{k} \sigma} c_{\mathbf{k} \sigma} 
+ \sum_{\mathbf{k} \sigma} \epsilon^{d}_{\mathbf{k}} d^{\dagger}_{\mathbf{k} \sigma} d_{\mathbf{k} \sigma} 
\\
\nonumber
& + V\sum_{ \mathbf{k} \sigma} \big( c^{\dagger}_{\mathbf{k}\sigma} d_{\mathbf{k}\sigma} + {\rm H.c.}\big) 
+ \Delta_{d}\sum_{ \mathbf{k} } \big( d^{\dagger}_{\mathbf{k}\uparrow} d^{\dagger}_{-\mathbf{k}\downarrow} + {\rm H.c.}\big)
\\
\nonumber
& + N_{\mathbf{Q}} U_{dc} \sum_{ \mathbf{k} \sigma}   \big( \delta_{d} c^{\dagger}_{\mathbf{k}\sigma} c_{\mathbf{k}+\mathbf{Q}\sigma} + \delta_{c}  d^{\dagger}_{\mathbf{k}\sigma} d_{\mathbf{k}+\mathbf{Q}\sigma} + {\rm H.c.}\big) 
\\
& - N\bigg( \frac{\Delta^{2}_{d}}{J_d} + U_{dc} n_d n_c + N_{\mathbf{Q}} U_{dc} \delta_d \delta_c  -\mu (n_c + n_d)\bigg).
\end{align}
where $\epk{}=-2t_c\left[\cos(k_x a)+\cos(k_y a)\right]$, $\epk{c}\equiv\epk{}+U_{dc}n^d-\mu$,
$\epk{d}\equiv\gamma\epk{}+U_{dc}n^c-\mu+\epsilon_{d0}$, and $\gamma=t_d/t_c$.
The latter is the inverse ratio of effective masses, while $\epsilon_{d0}$ plays the role of the relative depth between the centers of the bands. Finally, the order parameters are defined in the momentum space as
\begin{align}
\Delta_d &\equiv \frac{J_d}{N} \sum_{\vk} \med{\fk{-\vk\pu}\fk{\vk\up}} = \frac{J_d}{N} \sum_{\vk} \med{\fkm{\vk\up}\fkm{\vk\pu}},\\
\delta^d &\equiv \frac{1}{N}\sum_{\vk \sa}\left(\med{\fkm{\vk+\vQ\sa}\fk{\vk\sa}}+\med{\fkm{\vk\sa}\fk{\vk+\vQ\sa}}\right)~,\\
\delta^c &\equiv \frac{1}{N}\sum_{\vk \sa}\left(\med{\ckm{\vk+\vQ\sa}\ck{\vk\sa}}+\med{\ckm{\vk\sa}\ck{\vk+\vQ\sa}}\right)~.
\end{align}

We have not included interband pairing, since hybridization  already gives rise to hybrid pairs~\cite{Caldas2017}. We find however, that  explicitly including inter-band pairing in the Hamiltonian is detrimental to phase coexistence, as discussed below.

For $\mathbf{Q}=(\pi,\pi)$, one is able to block-diagonalize the Hamiltonian in Eq.\,\eqref{eq:momentumspaceHamil} using the Nambu's spinor basis
\begin{equation}\label{ref:basis}
    \Psi^{\dag}_{\mathbf{k}}\! =\! \left( c^{\dagger}_{\mathbf{k}\uparrow}, d^{\dagger}_{\mathbf{k}\uparrow}, c_{-\mathbf{k}\downarrow}, d_{-\mathbf{k}\downarrow}, c^{\dagger}_{\mathbf{k}+\mathbf{Q}\uparrow}, d^{\dagger}_{\mathbf{k}+\mathbf{Q}\uparrow}, c_{-\mathbf{k}-\mathbf{Q}\downarrow}, d_{-\mathbf{k}-\mathbf{Q}\downarrow}\right)~.
\end{equation}
That is, the MF Hamiltonian can be written in a quadratic form, $H_{MF}=\sum_{\vk}\Psi^{\dag}_{\vk}M\Psi_{\vk} + const.$, with the matrix representation
{\tiny
	\begin{align}\label{ref:M}
	& M= \notag \\ 
	& \begin{pmatrix}
	\epks{c}{} & V & 0 & 0 & U_{dc}\delta^{d} & 0  & 0 & 0   \\
	V & \epks{d}{} & 0 & \Delta_d & 0 & U_{dc}\delta^{c} & 0 & 0  \\
	0 & 0 & -\epkms{c}{} & -V & 0 & 0 & -U_{dc}\delta^{d} & 0 \\
	0 & \Delta_d & -V & -\epkms{d}{} & 0 & 0 & 0 & -U_{dc}\delta^{c} \\
	U_{dc}\delta^{d} & 0 & 0 & 0 & \epkqs{c}{} & V & 0 & 0 \\
	0 & U_{dc}\delta^{c} & 0 & 0 & V & \epkqs{d}{} & 0 & \Delta_d \\
	0 & 0 & -U_{dc}\delta^{d} & 0 & 0 & 0 & -\epkqms{c}{} & -V \\
	0 & 0 & 0 & -U_{dc}\delta^{c} & 0 & \Delta_d & -V & -\epkqms{d}{} 
	\end{pmatrix}
	\end{align}}
Here, the sums over $\mathbf{k}$'s are performed in the extended Brillouin zone (from $-\pi/a$ to $\pi/a$, in $k_x$ and $k_y$ directions), which makes the subspace $(\mathbf{k} \uparrow, -\mathbf{k} \downarrow, \mathbf{k} + \mathbf{Q} \uparrow,  -\mathbf{k} - \mathbf{Q} \downarrow )$ two-folded degenerated.
Similarly to the previous case, the free energy density is obtained from the eigenvalues $E_{m \mathbf{k}}$ of the matrix in Eq.\,\eqref{ref:M} as follows,
\begin{align}\label{free_energy}
F=-\frac{1}{2N}T\sum_{\vk}\sum_m\ln{[1+\exp{(-\beta E_{m\vk})}]} + const~,
\end{align}
with $m=1,\dots, 8$~\cite{Lopes2021}; the $1/2$ coefficient is due to the degeneracy.

Proceeding, we now turn to discuss the incommensurate case, which is challenging, since it usually requires large unit cells and, consequently, large blocks in momentum space.
Instead, here we employ a quasi-degenerate perturbation theory, i.e., we treat the charge modulation contribution to energy as a perturbation, correcting the nearly degenerate unperturbed eigenstates.
To this end, it is worth noticing that, in absence of charge modulation ($\delta_c = \delta_d = 0$) in Eq.\,\eqref{eq:momentumspaceHamil}, the Hamiltonian is non-folded degenerate block-diagonal in the subspace $(\mathbf{k} \uparrow, -\mathbf{k} \downarrow)$, which provide us the unperturbed eigenvalues $E^{0}_{n, \mathbf{k}}$ ($n=1,...,4$).

Within this strategy, we treat the fifth term on the right-hand side of Eq.\,\eqref{eq:momentumspaceHamil} as a perturbation, 
\begin{equation}\label{eq:perturbation}
\hat{V}(\mathbf{Q}) = U_{dc} \sum_{ \mathbf{k} \sigma}   \big( \delta_{d} c^{\dagger}_{\mathbf{k}\sigma} c_{\mathbf{k}+\mathbf{Q}\sigma} + \delta_{c}  d^{\dagger}_{\mathbf{k}\sigma} d_{\mathbf{k}+\mathbf{Q}\sigma} + {\rm H.c.}\big)~,
\end{equation}
fixing $N_{\mathbf{Q}}=1$, since it just renormalizes the parameters $\delta_c$ and $\delta_d$, but does not change the actual gap.
We recall that the corrections due to $\hat{V}(\mathbf{Q})$ are particularly relevant when $E^{0}_{n, \mathbf{k}} \approx E^{0}_{n^{'}, \mathbf{k} + \mathbf{Q}}$, lifting the degeneracy, while leaving the bands almost unchanged away from this point.
In view of this, one may span the Hamiltonian in $(\mathbf{k} \uparrow, -\mathbf{k} \downarrow)$ and $(\mathbf{k} + \mathbf{Q} \uparrow,  -\mathbf{k} - \mathbf{Q} \downarrow )$ subspaces, a procedure that leads to the same block represented in Eq.\,\eqref{ref:M}.
Similarly to the previous case, $\delta_c$ and $\delta_d$ are obtained by performing a self-consistent analysis to minimize the free energy, Eq.\,\eqref{free_energy}, with respect to all mean-field parameters, Eq.\,\eqref{self-consistent_eqs}; however we need to add a constant $-\frac{N}{2}\langle \hat{V}(\mathbf{Q}) \rangle = -N U_{dc} \delta_c \delta_d$\,\footnote{If one keeps $N_{\mathbf{Q}}$ in $\hat{V}(\mathbf{Q})$, the constant becomes $-\frac{N}{2}\langle \hat{V}(\mathbf{Q}) \rangle = -N N_{\mathbf{Q}}^{2} U_{dc} \delta_c \delta_d$, leading to the same result without $N_{\mathbf{Q}}$, but just renormalizing the values of $\delta$'s by $1/N_{\mathbf{Q}}$. Notice as well that the constant term from the perturbation theory is different from its corresponding one in Eq.\,\eqref{eq:momentumspaceHamil}, which is expected since the latter may disregard a few terms of the full Hamiltonian.}.

%%%%%%%%%%%%%%%%%%%%%%%%%%%%%%%%%%%%%%%%%%%%%%
\section{RESULTS}
\label{sec:results}
%%%%%%%%%%%%%%%%%%%%%%%%%%%%%%%%%%%%%%%%%%%%%%%

In what follows, we set the energy scale by the hopping of the $c$-orbitals $t_c=1.0$, while defining the lattice and Boltzmann constants as unities ($a=1.0$ and $k_B=1.0$). Unless otherwise explicitly mentioned, we assume a fixed ratio between the effective masses, $\gamma=0.4$. The latter is appropriate to describe the intermetallic compounds that we are particularly interested.

\begin{figure}[b]
    \centering
    \includegraphics[width=1.0\columnwidth]{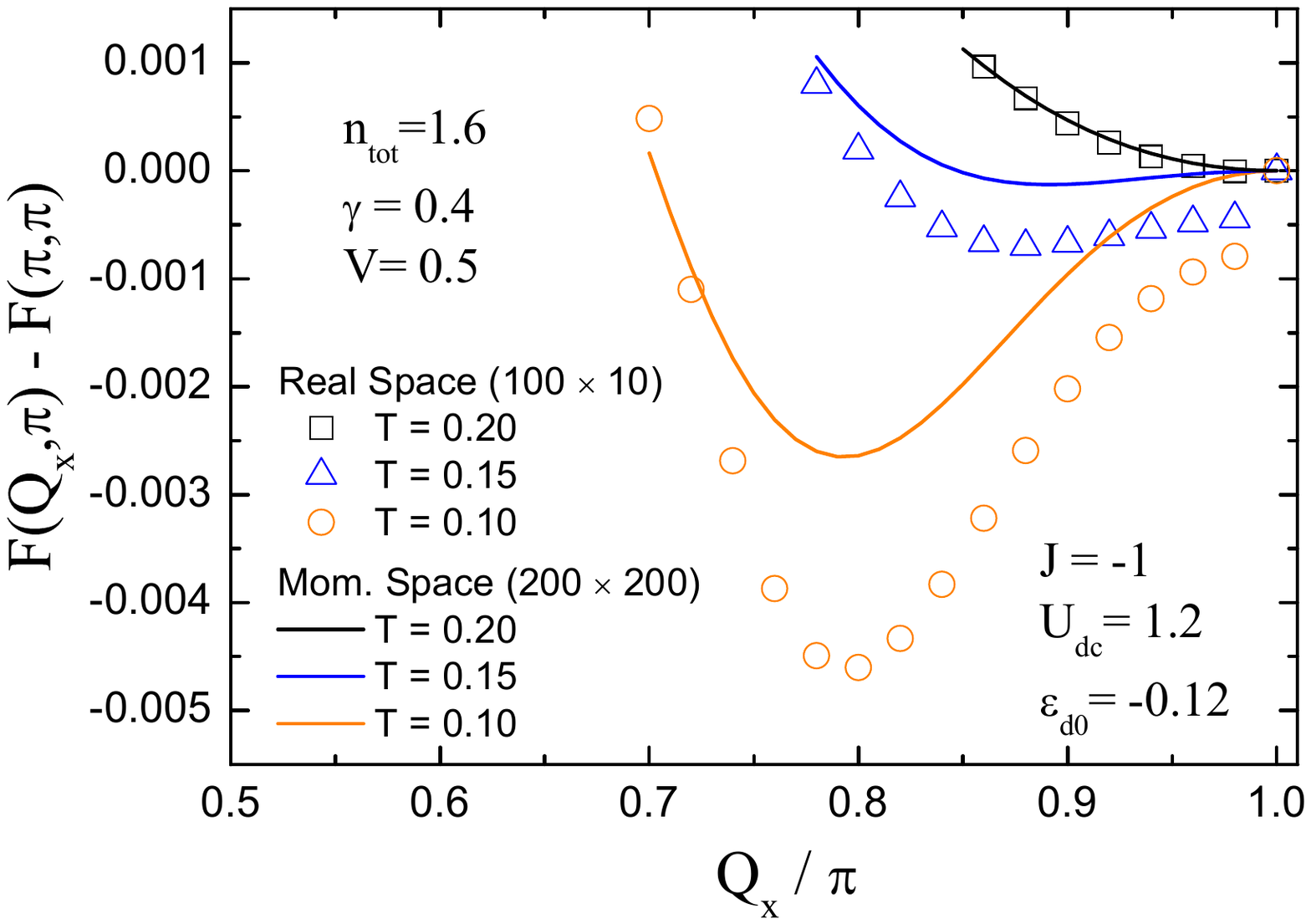}
    \caption{(Color online) Difference between the free energy for $\mathbf{Q}=(Q_x,\pi)$ and the staggered one, $\mathbf{Q}=(\pi,\pi)$, as a function of $q_x \equiv \mathbf{Q}_x/\pi$, for fixed parameters $\gamma=0.4$, $J=-1.0$, $U_{dc}=1.2$, $V = 0.5$, $\epsilon_{d0}=-0.12$, and $n_{\rm tot}=1.6$. The symbols correspond to the solutions for the full diagonalization at the real-space, for a $100 \times 10$ lattice, while the solid curves are those for the perturbation approach at the momentum-space, for a $200 \times 200$ system size.}
    \label{fig:fe_realsp01}
\end{figure}

We start discussing our results in the real-space Hamiltonian, Eq.\,\eqref{eq:realspaceHamil}, investigating the possible occurrence of inCDW order on these multiband intermetallic systems.
To this end, we examine a $100 \times 10$ lattice (i.e., 200 orbitals), varying $Q_x$, while keeping $Q_y = \pi$, which will be justified latter.
We recall that commensurate CDW is favored if the system is at the half-filling ($n_{tot} = 2.0$) due to nesting properties.
Therefore, in order to find out inCDW, one has to explore it for $n_{tot} < 2.0$. We also consider a difference in the single-particle $c$ and $d$-levels ($\epsilon_{d0} \neq 0.0$).
Given this, we first examine our system at $\ntot=1.6$ and $\epsilon_{d0}=-0.12$, whose results are displayed in Fig.\,\ref{fig:fe_realsp01}, which presents the difference of the free energy density as a function of $Q_x$ in comparison to the staggered case, for fixed $U_{dc} = 1.2$, $V = 0.5$, and $J = -1.0$.
As exhibited in Fig.\,\ref{fig:fe_realsp01} the commensurate CDW is favored at high temperatures, while an inCDW phase emerges when the temperature is reduced (notice the minima of the free energy density).
Interestingly, this result shows that the commensurate CDW phase is a very robust one, existing even far away from half-filling, and being stable at high temperatures.
On the other hand, the inCDW becomes more stable only at low temperatures, and in regions where the commensurate one is weakened (such as for $\ntot=1.6$), which indicates that this phase is less robust than the previous one. Indeed, this feature is present in most of our following results.

However, dealing with real-space problems is challenging, which demands hard numerical calculations even for small lattice sizes, as well as finite-size scaling analyses.
In view of this, hereafter we analyze the problem in the momentum-space, as discussed in subsection~\ref{subsec:mspace}.
As a first step towards this end, it is important to validate the perturbation theory approach by comparing its results with those of the real-space one.
Fig.~\ref{fig:fe_realsp01} presents this comparison, with the solid curves being the results of the perturbation theory approach, where is evident the quantitative and qualitative agreement between both methodologies.
It is also worth mentioning that, in the following results, we have performed a systematic analysis of the internal and Helmholtz energies as a function of the lattice size to avoid finite-size effects from the incommensurability of $\mathbf{Q}$.

\begin{figure}[t]
    \centering
    \includegraphics[width=1\columnwidth]{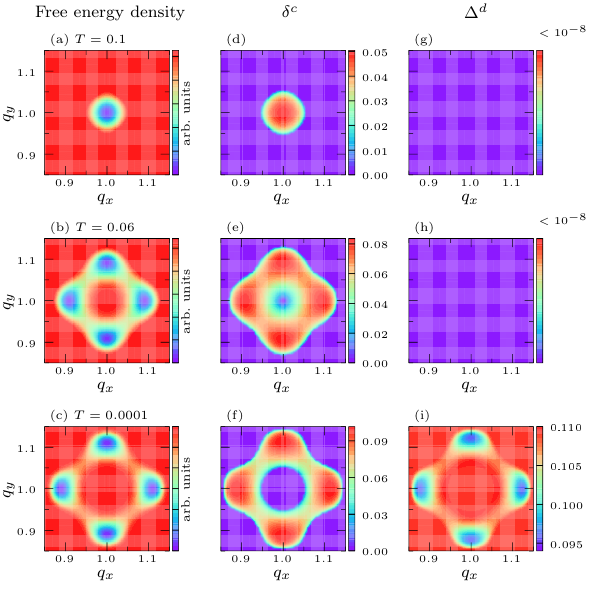}
    \caption{(Color online) Panels (a),~(b) and~(c) show maps of the free energy density, (d),~(e) and~(f) CDW order parameter $\delta^{c}$, and (g),~(h) and~(i) superconducting order parameter $\Delta^d$ as functions of $q_x$ and $q_y$, where $q_{x,y}=Q_{x,y}/\pi$. Each row represents a fixed value of temperature: $T$ = 0.1, $T$ = 0.06, and $T$ = 0.0001 from the top to the bottom. The parameters used are $\ntot=1.6$, $V=0.5$, $J=-1.0$, $\epsilon_{d0}=0.0$, and $U_{dc}=0.8$. As temperature decreases, the system goes from a commensurate CDW to an inCDW to a coexistence of inCDW and SC as can be inferred from the minima of the free energy density and the  values of the order parameters $\delta^c$ and $\Delta^d$ for different values of $q_x$ and $q_y$. The analysis always shows that the inCDW solution is obtained by a symmetric wave vector with $\vQ=(Q_x,\pi)\equiv(\pi,Q_y)$. In (g) and (h) the value of $\Delta^d$ is smaller than $10^{-8}$.}
    \label{fig:fe_vartemp}
\end{figure}

Given this, we now proceed within the momentum-space approach, determining the $\vQ$-vector that defines the inCDW or CDW order from the dependence of the free energy density on the components of the modulation wave vector $\vQ = ( Q_x,Q_y)$.
The panels (a)-(c) of Fig.~\ref{fig:fe_vartemp} show contour plots of the free energy densities as functions of the wave vector components for three selected temperatures, $T=0.1$, $0.06$ and $T=0.0001$ (in units of the hopping), while fixing  $\ntot=1.6$, $V=0.5$, $J=-1.0$, $\epsilon_{d0}=0.0$, and $U_{dc}=0.8$.
For $T=0.1$, the minimum of the free energy density occurs for a commensurate CDW state, while an inCDW emerges (with either $Q_x$ or $Q_y$  $\neq \pi$) at lower temperatures.
Notice that the latter breaks the $x$-$y$ symmetry, such that one may fix $Q_x = \pi$ and find $Q_y$ self-consistently, or vice versa. Here, we chose  to fix the component $Q_y=\pi$, while looking for different possibilities of $Q_x$.
We also present the behavior of the order parameters $\delta^{c}$ and $\Delta^d$ as functions of the normalized components $q_x$ and $q_y$ ($q_{x(y)}=Q_{x(y)}/\pi$) in the panels (d)-(f) and (g)-(i) of Fig.~\ref{fig:fe_vartemp}, respectively.
Interestingly, at $T=10^{-4}$, the minima of the free energy density leads to a coexistence between a superconducting and and inCDW phases [see, e.g., panels (c), (f), and (i)].

\begin{figure}[t]
    \centering
    \includegraphics[width=1\columnwidth]{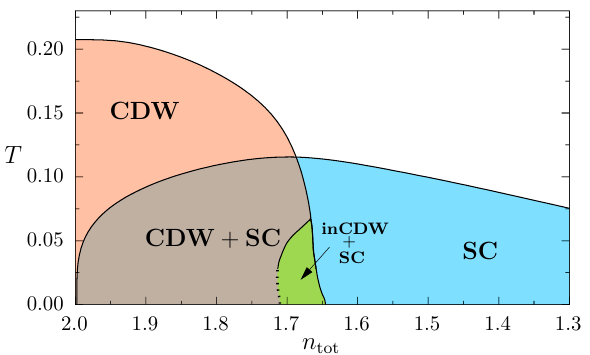}
    \caption{\label{fig:phdia_ntxtemp_v0_ed0} (Color online) The critical temperature of the inCDW, CDW and SC orders as a function of band filling ($\ntot$) for $J=-1.0$, $U_{dc}=0.8$, $V=0.0$, and $\epd=0.0$. One can see the small coexistence region between inCDW  and SC orders at very low temperatures and in between  the  CDW+SC phase and pure SC  phase. Continuous lines denote second-order phase transitions, while dotted lines indicate first-order ones.}
\end{figure}

By repeating the procedure outlined previously to other values of electronic densities and temperatures, one may obtain a phase diagram.
Fig.~\ref{fig:phdia_ntxtemp_v0_ed0} displays such a phase diagram for non-hybridized bands, with the same shift energy ($\epsilon_{d0}=0.0$), exhibiting the critical temperatures for inCDW, CDW, and SC orders.
Here and in what follows, the continuous lines denote second-order phase transitions, while the dotted lines indicate first-order ones\,\footnote{We systematically analyzed the free energy density as a function of the CDW and SC order parameters close to the transition points, taking the convergence criteria $\sim 10^{-8}$, to identify the order of the transitions.}.
As discussed above, the occurrence of a perfect nesting at the half-filling ($n_{\rm tot}=2.0$) favors the commensurate CDW state, and makes this phase extend for different occupations at higher temperatures.

The phase diagrams for such systems may be complex -- in particular, for the coexistence between charge ordering and superconductivity -- due to the strong dependence of the phases on the magnitude of the interactions.
For instance, by reducing $U_{dc}$, the SC phase may also appear at half-filling, coexisting with CDW~\cite{Lopes2021}.
But, here our focus is away from half-filling, due to unexpected behavior.
Going far away from half-filling is detrimental to the CDW phase, which is suppressed, leaving only SC at the ground state. Interestingly, at the boundary of the ``CDW+SC'' and the pure SC phases an inCDW one emerges, also coexisting with SC.
This behavior points out the fact that the charge-ordered phase is robust, with the system preferring changing its wave vector to accommodate the electrons into an inCDW order, instead of just destroying the CDW one.
However, the inCDW is less robust than the commensurate case, occurring for a small range of electronic density, and only for low temperatures; consistent with the results presented in Fig.~\ref{fig:fe_vartemp}.

 \begin{figure}[t]
    \centering
    \includegraphics[width=0.95\columnwidth]{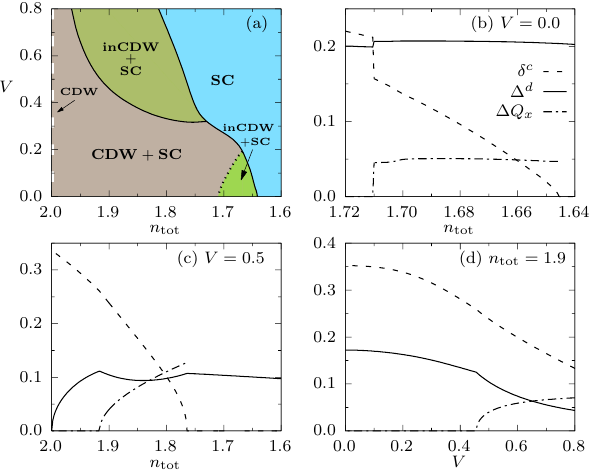}
    \caption{\label{fig:phdia_ntxvi_ufc08_ed0_t0001} (Color online) (a)~Phase diagram for fixed $T = 0.0001$, $J=-1.0$, $U_{dc}=0.8$, and $\epd=0.0$. (b),~(c) and (d)~order parameters as a function of $\ntot$ and $V$ for  $V = 0.0$, $V = 0.5$ and $\ntot = 1.9$, respectively. Note that two distinct and disconnected coexisting regions between inCDW and SC are observed on the edges of the CDW+SC and pure SC states. Observe that in~(a) we find first- (dotted lines) or second-order (continuous lines) phase transitions depending on the parameters.}
\end{figure}

In order to further investigate the features of the coexistence between inCDW and SC orders displayed in Fig.~\ref{fig:phdia_ntxtemp_v0_ed0}, we present a ``$V \times \ntot$'' phase diagram in Fig.~\ref{fig:phdia_ntxvi_ufc08_ed0_t0001}~(a), for fixed $T=0.0001$ (i.e.~at the ground state).
It shows two distinct and disconnected regions of coexistence between inCDW and SC on the edges with the pure SC phase, suggesting that two different processes may give rise to inCDW order, depending on the hybridization and/or the occupation number.
First, as displayed in Fig.~\ref{fig:phdia_ntxvi_ufc08_ed0_t0001}~(b) for $V=0$, the inCDW phase emerges presenting first-order transitions from CDW+SC to inCDW+SC, as noticed by the abrupt change in $\delta^c$ and a two minima behavior in the free energy density (not shown), while $\Delta Q_x=|1-Q_x/\pi|\approx0.05$.
Therefore, this small coexistence region seems to come from strong interactions at the neighborhood of the transition inCDW+SC to SC, avoiding the possibility of a reentrant behavior when we have no inCDW ordering.
On the other hand, for high values of $V$, $\Delta Q_x$ increases continuously as a function of $\ntot$, while the parameter $\delta^{c}$ goes to zero as $\ntot$ decreases, similar to a second-order phase transition.
$\Delta^d$ is exactly zero at the half-filling and has two kinks related to the finite value $\Delta Q_x$, one in the beginning and other in the end of inCDW phase, being less affected by the variation of $\ntot$, as presented in Fig.~\ref{fig:phdia_ntxvi_ufc08_ed0_t0001}~(c) for $V=0.5$.
This region is more robust and reflects the fact that hybridization between bands may induce inCDW phase at low temperatures. For $\ntot = 1.9$, one can see that $\delta^c$ and $\Delta^d$ goes to zero asymptotically, while $\Delta Q_x\neq 0$ appears for large values of $V$, as shown in Fig.~\ref{fig:phdia_ntxvi_ufc08_ed0_t0001}~(d). 

\begin{figure}[t]
    \centering
    \includegraphics[width=0.95\columnwidth]{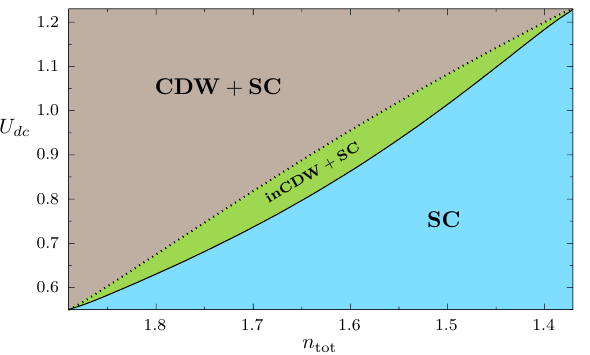}
    \caption{\label{fig:phdia_ntxudf_v0_ed0_t0001} (Color online) Phase diagram for $T=0.0001$ as a function of $U_{dc}$ and $\ntot$ for $J=-1.0$, $V=0.0$, and $\epd=0.0$. There is a narrow range exhibiting inCDW+SC, which separates CDW+SC from pure SC. The $\Delta Q_x$ component is almost constant and presents an abrupt variation as a function of $\ntot$, see Fig.~\ref{fig:phdia_ntxvi_ufc08_ed0_t0001}~(b). Note that the transition CDW+SC to inCDW+SC is first-order (dotted line) transition, while the inCDW+SC to pure SC is second-order one (continuous line).} 
\end{figure}

Similar conclusions are obtained when investigating the ``$U_{dc} \times \ntot$'' phase diagram, displayed in Fig.~\ref{fig:phdia_ntxudf_v0_ed0_t0001}.
Notice that the transition between the CDW+SC phase to the pure SC one always goes through a narrow inCDW+SC region before the system became purely SC. Here, $\Delta Q_x \lesssim 0.06$, producing an abrupt change in almost all phase diagram where $\Delta Q_x$ deviates from zero to a finite value.

\begin{figure}[t]
    \centering
    \includegraphics[width=0.95\columnwidth]{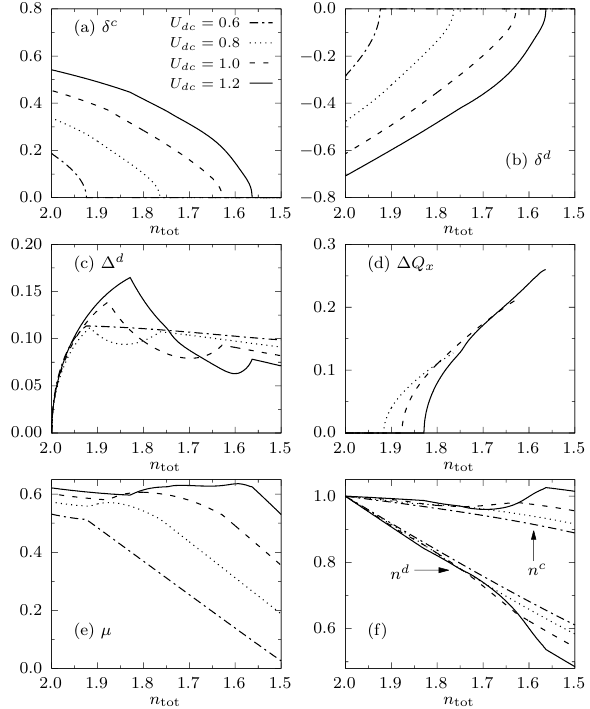}
    \caption{\label{fig:phdia_ntxtemp} (Color online) (a) $\delta^c$, (b) $\delta^d$, (c) $\Delta^d$, (d) $\Delta Q_x$, (e) $\mu$, (f) $n^d$ and $n^c$ as a function of $\ntot$ for different values of $U_{dc}$, and fixed $J=-1.0$, $V=0.5$, and $\epd=0.0$. The imbalance between $n^d$ and $n^c$ is detrimental to both CDW orders, i.e., CDW and inCDW, while SC is not directly affected by the electronic occupation in the different bands.}
\end{figure}

In Fig.~\ref{fig:phdia_ntxtemp} are depicted $\delta^c$, $\delta^d$, $\Delta^d$, $\Delta Q_x$, $\mu$, $n^d$ and $n^c$ as a function of $\ntot$ for different values of $U_{dc}$, and fixed $J=-1.0$, $V=0.5$, and $\epd=0.0$. One can see that the behavior of  $|\delta^{c}|$ and $|\delta^{d}|$ are very similar (see Figs.~\ref{fig:phdia_ntxtemp}~(a) and~(b)), which consequently justifies that fact that we have only shown the results for $\delta^c$ up to now. Fig.~\ref{fig:phdia_ntxtemp}~(c) shows the variation of $\Delta^d$ as we deviate from half-filling. Note that the SC order parameter initially increases and when $\Delta Q_x \neq 0 $, i.e., the inCDW state emerges, $\Delta^d$ changes its behavior, see Figs.~\ref{fig:phdia_ntxtemp}~(c) and ~(d). In other words, inCDW also exhibits an intrinsic competition with SC. Deviating further from half-filling,  $\delta^{c}$ and $\delta^{d}$ go to zero continuously, where $\Delta Q_x$ is defined only when $\delta^{c,d}\neq0$. $\Delta^d$ presents a monotonic decreasing and eventually $\Delta^d \rightarrow 0$ for very large deviations from $\ntot = 2.0$. In Fig.~\ref{fig:phdia_ntxtemp}~(e) we show that the chemical potential $\mu$ tends to decrease as a function of $\ntot$, with a slightly variation in the presence of  inCDW order, and it returns to decrease in the pure SC phase. Finally, Fig.~\ref{fig:phdia_ntxtemp}~(f) shows the imbalance between the number of electrons in different bands as a function of $\ntot$, which is detrimental to both CDW orders.

Thus, in general, CDW phase emerges at $\ntot=2.0$, where SC is suppressed. The coexistence between CDW and SC is obtained as the occupation number is slightly varied from half-filling. The SC reaches a maximum and coexistence between inCDW and SC takes place in the phase diagram below this point at low temperatures. As $\ntot$ keeps decreasing, only SC survives. Moreover, note that the inCDW emerges around  the fine-tuned density-driven CDW QCP. Therefore, we can conclude that the appearance of an inCDW order is intrinsically related to  quantum critical fluctuations, associated with the CDW  QCP (see also Fig.~\ref{fig:phdia_ntxtemp_v0_ed0}). From Fig.~\ref{fig:phdia_ntxtemp} it is clear that $U_{dc}$ also plays an important role on both CDW and inCDW phases since  a small $U_{dc}$ is sufficient to  suppress CDW as well as inCDW orders.

\begin{figure}[t]
    \centering
    \includegraphics[width=0.95\columnwidth]{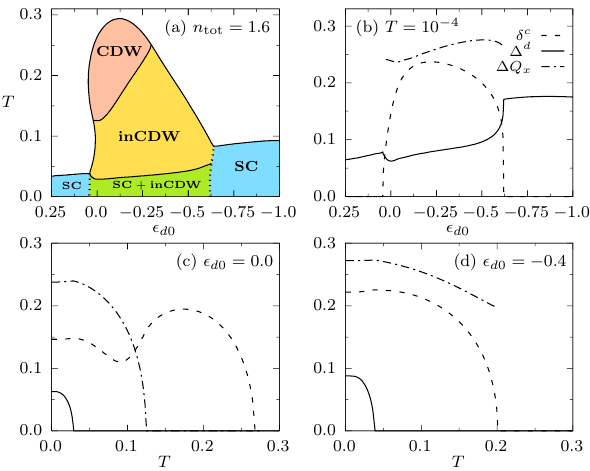}
    \caption{(Color online) (a) Critical temperatures as a function of $\epsilon_{d0}$ for fixed $U_{dc} = 1.2$, $V = 0.5$, $J = -1.0$ and $\ntot=1.6$. (b) Variation of $\delta^c$, $\Delta^d$ and $\Delta Q_x$ for a fixed $T = 10^{-4}$ as a function of $\epsilon_{d0}$. (c) Variation of $\delta^c$, $\Delta^d$ and $\Delta Q_x$ for a fixed $\epsilon_{d0} = 0.0$ as a function of $T$. (d) Order parameters as a function of $T$ for a fixed $\epsilon_{d0} = -0.4$. Note that $\Delta Q_x$ is defined only when $\delta^{c,d} \neq 0$. In~(a) Continuous line mean second-order phase transitions, while dotted lines correspond to first-order ones.}
    \label{fig:phase_diagnt16}
\end{figure}

So far, our results showing  coexistence between inCDW and SC orders were obtained for the case where the bands have the same center, i.e.,  $\epsilon_{d0} =0.0$. The effect of the shift between the centers of the bands $\epsilon_{d0} \neq 0.0$, is relevant experimentally when   doping  the $d$-bands with elements belonging to different rows of the periodic table as 3d, 4d, or 5d, but within the same column. Pressure also may affect  the relative positions of the bands.  Fig.~\ref{fig:phase_diagnt16}~(a) shows the critical temperatures as a function of $\epsilon_{d0}$ for fixed $U_{dc} = 1.2$, $V = 0.5$, $J = -1.0$ and $\ntot=1.6$. One can see a very rich phase diagram exhibiting multiple phases. For high $T$ and small $|\epsilon_{d0}|$, CDW order prevails. Cooling down the system, we obtain a robust pure inCDW phase. Further reducing $T$ we reach a coexistence region between inCDW and SC order. On the other hand, for large values of $|\epsilon_{d0}|$ both inCDW and CDW orders are suppressed and only SC survives. In Fig.~\ref{fig:phase_diagnt16}~(b) we present the variation of $\delta^c$, $\Delta^d$ and $\Delta Q_x$ for a fixed $T =0.0001$ as a function of $\epsilon_{d0}$. One can see that although $\Delta Q_x$ is different of zero and does not change considerably when it is defined, $\delta^c$ and $\Delta^d$ present two abrupt changes at the edges of the emergence of the inCDW order, which is an indication of a first-order phase transition. In addition, $\Delta^d$ is finite and almost constant at the pure SC states. Fig.~\ref{fig:phase_diagnt16}~(c) shows the variation of $\delta^c$, $\Delta^d$ and $\Delta Q_x$ for a fixed $\epsilon_{d0} = 0.0$ as a function of $T$. Note that all parameters are continuous and also observe that the inCDW state emerges at low temperatures changing the behavior of $\delta^c$, which indicates the transition to an inCDW state.
In Fig.~\ref{fig:phase_diagnt16}~(d) we present the order parameters as a function of $T$ for a fixed $\epsilon_{d0} = -0.4$. Note that both $\delta^c$ and $\Delta^d$ are continuous as a function of $T$, which suggests a second-order phase transition.

We emphasize that the imbalance between the number of electrons in different bands as a function of $\epsilon_{d0}$ (not shown) is detrimental to CDW orders, which is very similar to Fig.~\ref{fig:phdia_ntxtemp}~(f). Therefore, we can state that whether the CDW ordering commensurate or incommensurate, increasing $|\epsilon_{d0}|$ may suppress both inCDW and CDW states, while only SC remains weakly affected by relative band shifts, as shown in Fig.~\ref{fig:phase_diagnt16}~(a). 

We also have calculated the Fermi surfaces (FS) of both $c$- and $d$-bands in normal state ($\epsilon_{k}^{c,d} = 0.0$ contours in the Brillouin zone), i.e., with no SC or CDW orders for Fig.~\ref{fig:phdia_ntxtemp_v0_ed0} and Fig.~\ref{fig:phase_diagnt16}~(a) (not shown). Doing that, we obtain the expected FS structure for a square lattice depending on the band filling. Indeed, at half-filling ($n_{tot} = 2.0$) we find the nesting condition, i.e., the wave vector $\vec{Q}= (\pi,\pi)$ connects two points of the FS, as expected, which might favors the emergence of charge ordering at half-filling, as discussed previously. As we deviate from half-filling we have no longer the nesting condition, as expected too. One of the bands ($c$-band) remains close to the nesting condition while the another band ($d$-band) moves away from the nesting condition, which is detrimental to charge ordering. The latter is due to the parameter $\gamma$ that affect the $d$-band as well as the imbalance between electrons in the bands ($n^{c,d}$) as a function of $\ntot$, see also Fig.~\ref{fig:phdia_ntxtemp}~(f).  We also investigated the FS structure in the case away from half-filling at the normal state, i.e., for $n_{tot} = 1.6$ as a function of $\epsilon_{d0}$, and we obtain, that while one of the bands is close to nesting condition the another band is far away from that. Therefore, in this case the nesting condition is not achieved since we have an occupation number different from half-filling. It is important to point out that as we increase $\epsilon_{d0}$ (in modulus) we can invert the band that is close to the nesting condition, i.e., one of the bands will always be distant from the nesting condition, which, again, is prejudicial to both charge orderings (CDW and inCDW). This aspects might explain the tendency for the persistent SC order in Fig.~\ref{fig:phdia_ntxtemp_v0_ed0} and Fig.~\ref{fig:phase_diagnt16}~(a), while the charge ordering is suppressed as we moves away from half-filling or increase (in modulus) the relative shift between the bands.

\begin{figure}[t]
    \centering
    \includegraphics[width=0.95\columnwidth]{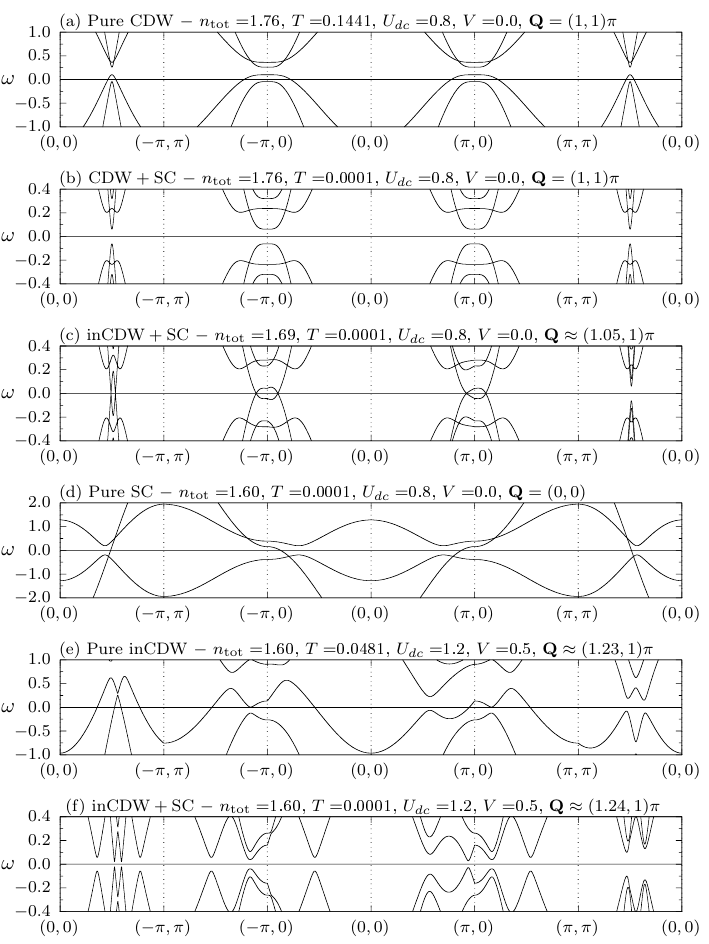}
    \caption{\label{fig:bands_analysis} (Color online) Spectra of quasi-particle excitations plotted in the first two quadrants of the extended Brillouin Zone, with zoom at the Fermi level ($\omega=0$), for some selected phase diagrams: (a), (b), (c), and (d) use the same parameters of Fig.~\ref{fig:phdia_ntxtemp_v0_ed0}, i.e., $U_{dc}=0.8$, $J=-1.0$, $V=0.0$, and $\epd=0.0$. (a)~bands for fixed $\ntot=1.76$ and $T=0.1441$, representing a pure CDW metallic phase, (b)~ in the CDW+SC region for $T=0.0001$ keeping $\ntot=1.76$,~(c) for the small inCDW+SC region, reached also by a first order transition,  taking fixed $\ntot=1.69$ and $T=0.0001$, and~(d) in the pure SC state with $\ntot=1.60$ keeping $T=0.0001$. On the other hand, the spectra in (e) and (f) use the same parameters of Fig.~\ref{fig:phase_diagnt16}~(a), that is, $\ntot=1.60$, $U_{dc}=1.2$, $J=-1.0$, and $V=0.5$.~(e) bands for the pure inCDW state for fixed $\epsilon_{d0}=0.0$ and $T = 0.0481$, and~(f) the spectra for inCDW+SC region with $T = 0.0001$ and $\epsilon_{d0}=0.0$.}
\end{figure}

In Fig.~\ref{fig:bands_analysis} we show the dispersion relations of the quasi-particle excitations with a zoom at the Fermi level ($\omega=0.0$),  for some selected points in the phase diagrams presented throughout the text. The figures are plotted in the first two quadrants of the extended Brillouin Zone.  In Figs.~\ref{fig:bands_analysis}~(a),~(b),~(c), and~(d) we use the same parameters of Fig.~\ref{fig:phdia_ntxtemp_v0_ed0}, i.e., $U_{dc}=0.8$, $J=-1.0$, $V=0.0$, and $\epd=0.0$. For $\ntot=1.76$ and $T=0.1441$, i.e., away from half-filling, the pure CDW state that appears at high $T$ exhibits a metallic aspect since, in this case, the bands of electronic excitations cross the Fermi level, as can be seen in Fig.~\ref{fig:bands_analysis}~(a). In Fig.~\ref{fig:bands_analysis}~(b) we display the dispersion relations when the system enters the CDW+SC region  reducing the temperature of the system to $T=0.0001$, and keeping $\ntot = 1.76$ fixed. Note that the spectrum of excitations now is completely gapped out along the Fermi surface when we have coexistence of phases. This can be understood from the fact that at the pure CDW state away from half-filling, there are available states at the Fermi level that might be responsible for the appearance of the additional SC state. In Fig.~\ref{fig:bands_analysis}~(c) we present the spectrum of excitations in the small coexistence region of inCDW+SC at $\ntot=1.69$ at low temperatures. Note that now the system is gapless around the points ($\pi,0$), (-$\pi,0$), and between (-$\pi,\pi$) and ($0,0$), at the Fermi surface, while between $(\pi,\pi)$ and $(0,0)$ it is gapped. The latter is a specific feature of the small inCDW+SC region that emerges as a function of $\ntot$. Here and afterwards, we choose to display only the values of $Q_x > \pi$ for the inCDW phase. The results for $Q_x<\pi$ and their respective negative values give rise to equivalent spectra as anticipated by the free energy density analysis made earlier, and will not be shown.  

In Fig.~\ref{fig:bands_analysis}~(d) we show the dispersions  for the pure SC state as the system  deviates further from half-filling. Note that the $d$-band is completely gapped out, while the $c$-band is not affected by the emergence of SC, as expected, since, for $V=0$,  the SC arises from an intraband attractive interaction. In Figs.~\ref{fig:bands_analysis}~(e), and (f), we analyze the spectra of excitations for the same parameters used in Fig.~\ref{fig:phase_diagnt16}~(a), i.e.,  $\ntot=1.60$, $U_{dc}=1.2$, $J=-1.0$, and $V=0.5$. Note that for the pure inCDW state, represented here by $\epsilon_{d0}=0.0$ and $T = 0.0481$, the system also exhibits a metallic character, whereupon multiple bands cross the Fermi level. Moreover, observe that at the pure inCDW state the peaks of the bands are no longer symmetrical and the coexistence of inCDW+SC obtained reducing the temperature of the system is again justified since there are remaining states at the Fermi level. These may lead to the emergence of the SC state also, i.e., the coexistence of inCDW+SC, gapping out the entire Fermi surface, as shown in Fig.~\ref{fig:bands_analysis}~(f) for $T=0.0001$. So, we can state that the spectra of excitations in the system can present a variety of behavior  depending on the parameters and phases of the model. At half-filling we confirm that the system is completely gapped out due to the nesting for the pure CDW state (not shown), as expected. The latter corroborates the fact that there is no coexistence of phases at half-filling in Fig.~\ref{fig:phdia_ntxtemp_v0_ed0}.

%%%%%%%%%%%%%%%%%%%%%%%%%%%%%%%%%%%%%%%%%%%%%%%%%
\section{Conclusions and remarks}
\label{sec:concl}
%%%%%%%%%%%%%%%%%%%%%%%%%%%%%%%%%%%%%%%%%%%%%%%%%%%

It is known~\cite{Lopes2021} that commensurate CDW and SC orders may emerge on multiband intermetallic systems and their alloys and that there is an intrinsic competition between these phases. However, there are regions in the space of parameters where these phases might coexist homogeneously. On the other hand, the coexistence of inCDW and SC phases has  been reported in several materials, such as, TMDs compounds~\cite{Sipos2008,Liu2013,Li2017,Gabovich2002,Wagner2008,Kogar2017,Chen2015,Wen2020,Song2022}, Ni- and Fe-based pnictides~\cite{Lee2019,Sefat2009}, and Y-, Bi-, and Hg-based  cuprates~\cite{Ghiringhelli2012,Chang2012,Achkar2012,Comin2014,daSilvaNeto2014,Wu2015,Tabis2014,Blackburn2013,LeTacon2014}. It is remarkable that these experimental results resemble those of the high $T_c$ cuprates, which in itself justifies its importance and relevance in understanding the coexistence of these  phases of matter in the phase diagram of these compounds.

In this work we have studied the possible appearance of an inCDW state on multiband intermetallic systems that present CDW and SC competing states. The systems we investigate are  two-dimensional and inhabit a square lattice. They are characterized by a $d$-band of moderately correlated electrons coexisting and a large conductance $c$-band. Our aim was to provide a deeper insight on this collective quantum states in solids and investigate how these phases depend on the parameters of the model. These parameters can, in principle, be externally controlled by doping and/or applying pressure in these systems. Although we do not wish to model any particular system, the phase diagrams we obtain show an overall agreement when compared to the multiband intermetallic systems we want to describe. This ascertains our model and the effects of the variation of its parameters in the phase diagrams. 

The most important and new aspect of the present study is the consideration of an incommensurate charge density wave vector that is present in real systems. In order to search for a possible  inCDW phase,  we allow the charge density modulation wave vector, in principle, to be completely arbitrary, given by $\vQ = (Q_x,Q_y)$. We minimize the free energy density numerically through the Hellmann-Feynman theorem which leads to self-consistent equations that allow to obtain the phase diagrams as  function of the several parameters of the model.

We treat the electronic correlation within a Hartree-Fock mean-field approximation, in both real and momentum space configuration, which has been shown adequate to describe systems,  such as the intermetallic compounds that we are interested. Also our BCS approach to the attractive interactions is consistent with the kind of superconductivity observed in the intermetallic compounds. 

The free energy density analysis shows that the inCDW state presents a charge ordering that breaking the $x$-$y$ symmetry, where the modulation wave vector that minimizes the free energy density is given by $\vQ = (Q_x,Q_y) = (\pi,Q_y) = (Q_x,\pi)$. We can use this  result to fix one of the components of $\vQ$ and obtain the phase diagrams of the model including an inCDW phase. We investigate how parameters, such as, band filling, temperature, hybridization, strength of inCDW/CDW interaction and the relative depth between the bands affect the phase diagram and yield the possibility of an inCDW phase. From the free energy density analysis we also identify the order of the transitions depending on the parameters of the model.

We show that varying the total occupation number of the bands and hybridization, we obtain a coexistence region of inCDW and SC at low temperatures,  close to the coexistence of CDW and SC and pure SC orders. The relative depth between bands ($\epsilon_{d0}$) can be tuned to give rise to a robust inCDW state for small $T$. Moreover,  our results convincingly show that increasing  the relative depth between bands is detrimental to both inCDW and CDW states.  Both phases are very sensitive to  the imbalance between electrons in different bands. By contrast, the SC state that appears in these intermetallic systems is not much affected by varying $\epsilon_{d0}$.

From the analysis of the spectra of excitations, we show that the possibility for coexistence of phases, depending on the parameters of the model, is intrinsically related to the metallic aspect of the CDW and inCDW states away from half-filling. In this case, both charge orders, CDW and inCDW, leave  states available at the Fermi level that might be responsible for the emergence of the additional SC state, giving rise to the coexistence of phases. At the coexistence of phases the system is completely gapped out along the Fermi surface, except for the small coexistence inCDW+SC region that appears as a function of $\ntot$ in Fig.~\ref{fig:phdia_ntxtemp_v0_ed0}. The latter exhibits a gapless spectrum around some points at the Fermi surface. In the case of $V=0$, the pure SC opens a complete gap in $d$-band only since SC is due to an intraband interaction in this band only. We also confirm that at half-filling the system is completely gapped out in pure CDW, due to nesting. This explains the fact that there are no coexistence of phases at half-filling in Fig.~\ref{fig:phdia_ntxtemp_v0_ed0}.

In addition, our phase diagrams for CDW, inCDW and SC orders as a function of occupation number endorse the results obtained from a different approach considering a phononic origin to the
CDW/inCDW orderings in a single-band model\,\cite{Dee19}. In this sense, we can also state that our results corroborate the fact that we cannot distinguish between electronic or phononic driven transitions in a mean-field approximation.

%%%%%%%%%%%%%%%%%%%%%%%%%%%%%%%%%%%%%%%%%%%%%%%%%%%%%%%%%%%%%%%
\section*{ACKNOWLEDGMENTS}
%%%%%%%%%%%%%%%%%%%%%%%%%%%%%%%%%%%%%%%%%%%%%%%%%%%%%%%%%%%%%%%

We would like to thank the Brazilian agencies \textit{Funda\c{c}\~ao Carlos Chagas Filho de Amparo \`a Pesquisa do Estado do
Rio de Janeiro} (FAPERJ), \textit{Coordena\c{c}\~ao de Aperfeiçoamento de Pessoal de N\'{\i}vel Superior} (CAPES) and \textit{Conselho Nacional de Desenvolvimento Cient\'{\i}fico e Tecnol\'{o}gico} (CNPq) for partial financial support. N.L. would like to thank the FAPERJ for the postdoctoral fellowship of the \textit{Programa de P\'{o}s-Doutorado Nota 10 - 2020} (E-26/202.184/2020) as well as for the \textit{Bolsa de Bancada para Projetos} (E-26/202.185/2020). D.R. would like to thank financial support from the Peruvian Agency CONCYTEC, grant number PE501079382-2022-PROCIENCIA. N.C.C.~acknowledges financial support from CNPq, Grant No. 313065/2021-7, and from FAPERJ, Grant No.\,200.258/2023 (SEI-260003/000623/2023).  C.T.~acknowledges partial  support provided by the Brazilian-France Agreement CAPES-COFECUB (No.~88881.192345/2018-01). Finally, we would like to thank the COTEC (CBPF) for making available their facilities for
the numerical calculations on the \textit{Cluster HPC}.

%%%%%%%%%%%%%%%%%%%%%%%%%%%%%%%%%%%%%%%%%%%%%%%%%%%%%%%%%%%%%%%%%%
\bibliographystyle{apsrev4-1}
\bibliography{biblio_incdw_220304}

\end{document}